\documentstyle[buckow]{article}

\newcommand{\cL}{{\cal L}}
\newcommand{\cM}{{\cal M}}

\newcommand{\cR}{{\cal R}}

\newcommand{\e}{e} 
\newcommand{\be}{\begin{equation}}
\newcommand{\ee}{\end{equation}}
\newcommand{\bea}{\begin{eqnarray}}
\newcommand{\eea}{\end{eqnarray}}
\newcommand{\bi}{\begin{itemize}}
\newcommand{\ei}{\end{itemize}}
\newcommand{\bt}{\begin{tabular}}
\newcommand{\et}{\end{tabular}}
\newcommand{\bc}{\begin{center}}
\newcommand{\ec}{\end{center}}
\def\aa{\hat\chi}

\begin {document}
\hfill hep-th/9901137

\hfill CPT-99/P.3753

\def\titleline{Quantum Corrections  
in the 
Hypermultiplet Moduli Space}
\def\authors{ Holger G\"unther\1ad, Carl Herrmann\2ad,
Jan Louis\1ad}
\def\addresses{
\1ad{Martin--Luther--Universit\"at 
Halle--Wittenberg,\\
FB Physik, D-06099 Halle, Germany}\\
\2ad{Centre de Physique Th\'eorique,
CNRS -- Luminy Case 907,\\
F-13288 Marseille Cedex 9, France}
}
\def\abstracttext{
Quantum corrections in the hypermultiplet moduli 
space of type IIA string theories compactified on 
Calabi-Yau threefolds are investigated.}
\large
\makefront

\begin{center}
{\it Talk presented by J.\ Louis at the 32nd Symposium
 Ahrenshoop on the Theory of Elementary Particles,
Buckow, Germany, September 1 - 5, 1998.}
\end{center}

\vskip 20pt
String theories with $N=2$ supersymmetry in 
four space-time
dimensions ($d=4$) can be constructed either
by compactifying the type II string on a 
Calabi--Yau
threefold $Y_3$ or the heterotic string on
$K3\times T^2 $.
It is believed that the resulting string vacua all
reside in the same moduli space and that
any given string vacuum generically
has 2 dual descriptions:
either as a type II vacuum or as a heterotic vacuum
\cite{kv,fhsv}.
The spectrum and the low energy effective theory
are strongly constrained by $N=2$ supersymmetry
but also 
depend on the
specific `data' of the Calabi--Yau compactification.
The massless states come in $N=2$
multiplets which are either
vector multiplets $V$ containing a complex scalar,
a vector and two Weyl fermions or
hypermultiplets $H$ which contain four real
scalars and two Weyl fermions.
Apart from these two `standard' $N=2$ multiplets
there also exist two further multiplets containing
an antisymmetric tensor $B_{\mu\nu}$:
the tensor multiplet \cite{dwvh} 
containing
three real scalars, one  $B_{\mu\nu}$
 and two Weyl fermions
as well as the vector tensor multiplet 
containing one real scalar, a vector, one $B_{\mu\nu}$
 and two Weyl fermions \cite{SSW,DKLL,TVT}.
In $d=4$ an antisymmetric tensor
is dual to a scalar and thus the latter multiplets 
can be dualized to  a
hyper- or vector multiplet, respectively.

In type IIA string vacua one has
$h_{1,1}$ vector multiplets,
$h_{1,2}$ hypermultiplets
and one tensor multiplet \cite{LF}.
The Hodge numbers 
$h_{1,1}$ and $h_{1,2}$ count the non-trivial 
$(1,1)$ and $(1,2)$ forms on $Y_3$
while the additional tensor multiplet
is universal and contains the 
type IIA dilaton.\footnote{
In type  IIB vacua 
one finds $h_{1,2}$ vector multiplets,
$h_{1,1}$ tensor multiplets
and one universal multiplet
which contains two antisymmetric tensors.
So far there is no off-shell formulation of this
multiplet known but it is likely to exist.
In any case one can always dualize one of the antisymmetric tensors and
obtain
an additional, universal tensor multiplet containing
the type IIB dilaton.}
%
In heterotic string compactifications
the situation is slightly more involved.
The dilaton sits in a vector-tensor multiplet,
the moduli of the $K3$ form hypermultiplets
while the moduli of $T^2$ come in vector multiplets.
In addition there are generically moduli
which arise from the gauge bundle and they can
be part of either vector- or hypermultiplets.

$N=2$ supergravity severely constrains the interactions
among these multiplets.  In particular, the complex
scalars of the vector multiplets are coordinates on a
special K\"ahler manifold $\cM_V$ \cite{wp} while the
real scalars of the hypermultiplets are coordinates on
a quaternionic manifold $\cM_H$ \cite{bw}.  Locally the
two spaces form a direct product \cite{wlp}, i.e.
\be
\label{modulispace} 
\cM=\cM_V\otimes \cM_H \ .  
\ee

In string compactifications
both components obey a 
non-renormalization theorem \cite{DKLL,strom}.
In heterotic vacua 
the dilaton is part of a  
vector-tensor multiplet  or  its 
dual  vector multiplet. 
The fact that the dilaton organizes 
the string perturbation theory
together with the product structure of the moduli 
space (\ref{modulispace}) implies
that the moduli space of the hypermultiplets 
is determined at the string tree level and 
receives no further perturbative or non-perturbative
corrections, i.e. 
\be
\cM_H^{\rm Het}=\cM_H^{\rm (0)Het}\ .
\ee

In type IIA compactifications
the dilaton resides in a tensor multiplet
or its dual hypermultiplet
and thus  the moduli space of the vector 
multiplets
is exact and not corrected either perturbatively
or non-perturbatively
\be
\cM_V^{\rm II} = \cM_V^{\rm (0) II}\ .
\ee

The conjectured duality between type IIA and heterotic 
$N=2$ string vacua implies that
the low energy effective theories 
have to be identical when all quantum corrections are
taken into account.
In particular their moduli spaces have to  coincide,
i.e.
\be
\cM_V^{\rm Het}=\cM_V^{\rm II}\ =\ \cM_V^{\rm (0)II}\ ,\qquad
\cM_H^{\rm Het}=\cM_H^{\rm II}\ =\ \cM_H^{\rm (0)Het}\ .
\ee
Thus, the entire  $\cM_V^{\rm Het}$ can be obtained
by doing a tree level computation 
in type IIA while the entire $\cM_H^{\rm II}$
can be obtained
by doing a tree level computation 
in the heterotic string.

So far the geometry and duality properties of $\cM_V$
have been extensively studied.  However, $\cM_H$ has
been much less studied
\cite{BS} - \cite{BB}  
and in particular
the duality conjecture has only been verified in a very
specific (and simple) example \cite{fhsv}.  In this
talk we study $\cM_H$ in string perturbation theory of
type IIA vacua.\footnote{Some specific 
non-perturbative
corrections in $\cM_H$ are dicussed in refs.\ 
\cite{SW,OV,BB}.}
This is a necessary first step in
order to further establish the type IIA -- heterotic
string duality.
  Apart from this aspect it is an
interesting question in its own right and furthermore
might also teach us more about the $N=1$ string
perturbation theory.

At the string tree level $\cM_H^{\rm (0)II}$ is not the
most general quaternionic manifold but constrained by
the c-map \cite{cfg}. 
The $h_{1,2}$ complex scalars of the NS-NS sector by
themselves span a special K\"ahler manifold 
which is characterized by a holomorphic
prepotential \cite{wp}.
In type IIA string vacua 
they pair up with $2\times h_{1,2}$
real scalars from the R-R sector 
to form hypermultiplets;
the  combined geometry of these $4\times h_{1,2}$
scalars is quaternionic.

In type IIA string perturbation theory we do not 
apriori know if
the c-map is preserved.  However, we do know the
following generic facts \cite{stro}:

\bi
\item
$N=2$ supersymmetry is unbroken in perturbation
theory and thus the geometry of ${\cal M}_H$
must be quaternionic.
\item
The
dilaton $\phi$ resides in a tensor multiplet 
together with $B_{\mu\nu}$ and two real 
scalars $\xi_0, \tilde\xi_0$ from the
R-R sector;
$\phi$ organizes the string perturbation theory.
\item
There are $h_{1,2}$ hypermultiplets
containing the scalars
$(Z^a, \bar Z^{\bar a}, \xi_a, \tilde\xi_a)$
where $a=1,\ldots,h_{1,2}$.
The complex $Z^a$ arise in the NS-NS sector
and  span a special K\"ahler manifold
at tree level. 
The $\xi_0, \tilde\xi_0,\xi_a, \tilde\xi_a$
arise in the R-R sector and thus enjoy
a   continuous
Peccei-Quinn (PQ) symmetry in perturbation theory
\be\label{PQ}
\xi_i \to \xi_i + \gamma_i\  , \qquad
\tilde\xi_i \to\tilde\xi_i +\tilde \gamma_i\
,\qquad \gamma_i,\tilde \gamma_i \in {\bf R}\ ,
\quad   i=0,\ldots,  h_{1,2}\ .
\ee

\item
The scalars $\xi_i, \tilde\xi_i$ always
appear in pairs in string amplitudes.

\item
In the large volume limit there is an additional
continuous PQ symmetry which acts on 
$Z^a - \bar Z^{\bar a}$ according to
\be\label{PQp}
Z^a - \bar Z^{\bar a} \to 
Z^a - \bar Z^{\bar a}
 + \hat\gamma^a, \qquad \hat\gamma^a \in {\bf R}\ ,
\ee
and also transforms the $\xi_i, \tilde\xi_i$
in a way specified in ref.~\cite{DVV}.
\ei
These features 
strongly constrain the perturbation
theory and do lead in fact
to a further (perturbative)
non-renormalization theorem.

Let us first study the simple case of 
a Calabi--Yau compactification
with $h_{(1,2)}=0$ which implies that 
only the dilaton multiplet (and $h_{(1,1)}$
vector multiplets)
are present.
The tree level Lagrangian (in the string frame)
for the dilaton multiplet reads \cite{ferrara,BCF}
\be
\e^{-1}\cL^{(0)} = 
e^{-2\phi}\left(
-\frac12\,\cR +
2\,(\partial_\mu \phi)^{2}
 -\frac{1}{6}\,(H_{\mu\nu\rho})^{2}\right)
-\partial^\mu
C\,\partial_\mu \bar{C} -
 H^{\mu}\,(C\partial_{\mu}\bar{C}-\bar{C}\partial_{\mu}C)\ ,
\ee
where we defined
$H_{\mu\nu\rho}=\partial_{[\mu} B_{\nu\rho]},\
H^{\lambda}=\frac{1}{6\e}
\epsilon^{\mu\nu\rho\lambda}H_{\mu\nu\rho},\
C= \xi_0+i \tilde\xi_0$
and we omitted the couplings of the 
vector multiplets.
In type IIA vacua there is a one-loop correction
to the Ricci scalar \cite{KK,AFMN}
given by
\be
\e^{-1}({\cal L}^{(0)}  + {\cal L}^{(1)})
\ =\  -\frac12\ (e^{-2\phi}+\aa)\, {\cal R} + \ldots\ ,
\ee
where $\aa$ is up to normalization factors
the Euler number $\chi$ of the Calabi--Yau
$\aa \sim \chi = 2(h_{1,1}-h_{1,2})$.
$N=2$ supersymmetry  uniquely determines
the one-loop correction of the dilaton
multiplet to be
\bea\label{Hmetric}
\e^{-1}(\cL^{(0)} + {\cal L}^{(1)})  &=&  
(e^{-2\phi}+\aa) \left(
-\frac12\,\cR -\frac{1}{6}\,(H_{\mu\nu\rho})^{2}\right)
+2\, (e^{-2\phi}+\aa)^{-1}e^{-4\phi}\,
(\partial_\mu \phi)^{2}\nonumber\\ 
&&-\, \partial^\mu
C\,\partial_\mu \bar{C} -
 H^{\mu}\,(C\partial_{\mu}\bar{C}-\bar{C}\partial_{\mu}C)\ .
\eea
This action can be put into a more familiar
form by dualizing $B_{\mu\nu}$ to a
scalar field $a$ and perform a Weyl rescaling
to the Einstein frame. One obtains
\be\label{Smetric}
\e^{-1}\cL \ = \ -\frac12\,\cR
-{\left| \partial_\mu S -2 \bar{C}\partial_\mu C
\right|^{2}\over
(S+\bar S + 2\aa -2 C\bar C)^2}
- {2 \left|\partial_\mu C\right|^2 \over
(S+\bar S + 2\aa -2 C\bar C)}\ ,
\ee
where
$S  =  e^{-2\phi}+i\, a + C\bar C$.
The metric of the scalar fields is a K\"ahler
metric with a K\"ahler potential
\be
K= - \ln(S+\bar S +2\aa -2 C\bar C).
\ee

For $\aa = 0$ one recovers the well known 
tree level manifold $SU(2,1)/U(2)$ \cite{cfg,ferrara}.
For $\aa \neq 0$ we obtain precisely the metric 
conjectured by Strominger \cite{stro}.
Expanding the metric of eq.~(\ref{Smetric})
around large $S$ (weak coupling) it appears
to have contributions at all orders in perturbation theory.
However, as we just showed this is an artefact
of the dualization and the definition 
of the $S$ field.
In the field basis where the antisymmetric
tensor is used (which is the appropriate basis of 
string perturbation theory)
this correction is manifestly one-loop.\footnote{This
solves the apparent puzzle in \cite{stro}
where compactification of $R^4$ terms
in M-theory \cite{GG,GV} were used 
to identify the one-loop correction in the $S$-basis. The resulting
metric was not quaternionic which lead Strominger
to conjecture his ``all-loop'' formula.
However, since $N=2$ supersymmetry 
is unbroken one does
expect a quaternionic metric at each loop order.
What we just showed is that Stromingers 
all-loop formula is in fact only one-loop
in the appropriate string basis.}

Furthermore, one can show that 
there is a perturbative non-renormalization theorem
in that
the action of eqs.~(\ref{Hmetric}), (\ref{Smetric})
is exact in perturbation theory 
and does not receive 
any  perturbative corrections beyond one-loop
\cite{GHL}.\footnote{Of course,
non-perturbative corrections do appear \cite{SW,OV,BB}.}
This can be seen from the last term in
(\ref{Hmetric})
which cannot be multiplied by any power
of $e^{-2\phi}$ (as would be necessary for higher
loop corrections) without violating the
Peccei-Quinn symmetry of eq.~(\ref{PQ}).\footnote{
In some sense this is a four-dimensional version
of the non-renormalization theorem discussed
in  \cite{GV}.}
Alternatively one can show that 
the action (\ref{Smetric})
is the unique action compatible with the 
quaternionic geometry and all  other 
perturbative properties enumerated above
\cite{GHL}.

Let us now come to the general case where
in addition to the dilaton multiplet also
$h_{(1,2)}$  hypermultiplets are present.
This situation is presently under investigation
and we only indicate our preliminary results here \cite{GHL}.
The known loop corrections are \cite{KK,AFMN}
\bea
e^{-1}\cL &=&  -\frac12\ (e^{-2\phi}+\aa)\, {\cal R} 
-\,(e^{-2\phi}-\aa)\ G_{a\bar b} \partial^\mu Z^a \partial_\mu \bar
Z^{\bar b}\nonumber\\
&&+\frac14\,
 H^{\mu}\, \left(\tilde\xi^i \partial_\mu \xi_i -
    \xi^i \partial_\mu \tilde\xi_i  -2 \aa  V_\mu \right)
+\ldots
\eea
where $G_{a\bar b}$ is the tree level
K\"ahler metric
of the $Z^a$
and 
$V_\mu = K_a\partial_\mu Z^a -  K_{\bar a}
\partial_\mu \bar Z^{\bar a}$.
Thus, the loop correction is universal
in that there is a universal correction to $G_{a\bar b}$
and the 
Calabi-Yau geometry is not modified.
Furthermore, the presence of the $H^{\mu}$
coupling together with the perturbative
PQ-symmetries seem to imply 
a non-renormalization theorem:
The 
geometry of ${\cal M}_H$   has
a universal  correction at one-loop
but no higher perturbative corrections.
More details will be presented in \cite{GHL}.

\vskip0.5cm
\noindent
{\large \bf Acknowledgements}\\
This work is supported in part by:
GIF -- the German--Israeli
Foundation for Scientific Research (J.L.),
the DAAD -- the German Academic Exchange Service
(C.H.) and the Landesgraduiertenf\"orderung
Sachsen-Anhalt (H.G.).

We thank I.\ Antoniadis and  B.\ de Wit for 
usefull conversations.
C.H.\ 
thanks Jan Louis and his group for
their warm hospitality during a stay in Halle 
where this
collaboration was initiated.
J.L.\ thanks the organizers of the conference
for providing such a pleasant and 
stimulating atmosphere.

\end{document}